\documentstyle[preprint,revtex]{aps}
\tightenlines
\begin{document}
\draft

\preprint{SLAC--PUB--6017}
\medskip
\preprint{WIS--92/99/Dec--PH}
\medskip
\preprint{December 1992}
\medskip
\preprint{T/E}

\begin{title}
The Subleading Isgur-Wise Form Factor $\chi_3(v\!\cdot\!v')$\\
to Order $\alpha_s$ in QCD Sum Rules
\end{title}

\author{Matthias Neubert}
\begin{instit}
Stanford Linear Accelerator Center\\
Stanford University, Stanford, California 94309
\end{instit}

\author{Zoltan Ligeti and Yosef Nir}
\begin{instit}
Weizmann Institute of Science\\
Physics Department, Rehovot 76100, Israel
\end{instit}


\begin{abstract}
We calculate the contributions arising at order $\alpha_s$ in the QCD
sum rule for the spin-symmetry violating universal function
$\chi_3(v\cdot v')$, which appears at order $1/m_Q$ in the heavy quark
expansion of meson form factors. In particular, we derive the two-loop
perturbative contribution to the sum rule. Over the kinematic range
accessible in $B\to D^{(*)}\ell\,\nu$ decays, we find that
$\chi_3(v\cdot v')$ does not exceed the level of $\sim 1\%$, indicating
that power corrections induced by the chromo-magnetic operator in the
heavy quark expansion are small.
\end{abstract}


\centerline{(submitted to Physical Review D)}
\newpage

\section{Introduction}

In the heavy quark effective theory (HQET), the hadronic matrix
elements describing the semileptonic decays $M(v)\to M'(v')\,
\ell\,\nu$, where $M$ and $M'$ are pseudoscalar or vector mesons
containing a heavy quark, can be systematically expanded in inverse
powers of the heavy quark masses \cite{Eich,Grin,Geor,Mann,Falk}. The
coefficients in this expansion are $m_Q$-independent, universal
functions of the kinematic variable $y=v\cdot v'$. These so-called
Isgur-Wise form factors characterize the properties of the cloud of
light quarks and gluons surrounding the heavy quarks, which act as
static color sources. At leading order, a single function $\xi(y)$
suffices to parameterize all matrix elements \cite{Isgu}. This is
expressed in the compact trace formula \cite{Falk,Bjor}
\begin{equation}
   \langle M'(v') |\,J(0)\,| M(v) \rangle = - \xi(y)\,{\rm tr}
   \big\{\,\overline{\cal M}'(v')\,\Gamma\,{\cal M}(v)\big\} \,,
\end{equation}
which is a consequence of a spin-flavor symmetry that QCD reveals in
the limit $m_Q\to\infty$ \cite{Isgu,Volo}. Here $J=\bar h'(v')\,
\Gamma\,h(v)$ is an arbitrary heavy quark current, $h(v)$ describes a
heavy quark with velocity $v$, and
\begin{equation}
   {\cal{M}}(v) = \sqrt{m_M}\,P_+\,
   \cases{ -\gamma_5 &; pseudoscalar meson \cr
           \rlap/\epsilon &; vector meson \cr}
\end{equation}
is a spin wave function that describes correctly the transformation
properties (under boosts and heavy quark spin rotations) of the meson
states in the effective theory. $P_+=\case{1}/{2}(1+\rlap/v)$ is an
on-shell projection operator, and $\epsilon$ denotes the polarization
of the vector meson. Vector current conservation implies that the
Isgur-Wise function $\xi(y)$ is normalized to unity at zero recoil,
where the velocities of the initial and final meson are the same, and
the maximum momentum is transferred to the lepton pair \cite{Isgu}.

At order $1/m_Q$, the spin-flavor symmetry is explicitly broken by the
presence of higher dimension operators in both the effective currents
and the effective Lagrangian of HQET. Some of these power corrections,
which violate the spin symmetry, arise from the chromo-magnetic
interaction between the spin of the heavy quark and the gluon field.
The corresponding term in the effective Lagrangian is \cite{FGL}
\begin{equation}\label{Lchromo}
   {\cal{L}}_{\rm mag} = {Z\over 2 m_Q}\,{\cal O}_{\rm mag} ~,~~
   {\cal O}_{\rm mag} = {g_s\over 2}\,
   \bar h(v)\sigma_{\alpha\beta} G^{\alpha\beta} h(v) \,,
\end{equation}
where $Z$ is a renormalization factor. The effects of an insertion of
this operator can be parameterized by a tensor form factor
$\chi^{\alpha\beta}$ \cite{Luke}:
\begin{equation}\label{chidef}
   \langle M'(v') |\,i\!\int\!{\rm d}z\,
   T\big\{J(0),{\cal O}_{\rm mag}(z)\big\}\,| M(v) \rangle
   = - \bar\Lambda\,{\rm tr}\Big\{\chi^{\alpha\beta}(v,v')\,
   \overline{\cal{M}}'(v')\,\Gamma\,P_+\,i\sigma_{\alpha\beta}
   {\cal{M}}(v)\Big\} \,.
\end{equation}
The mass parameter $\bar\Lambda$ sets the canonical scale for power
corrections in HQET. In the $m_Q\to\infty$ limit, it measures the
finite mass difference between a heavy meson and the heavy quark that
it contains \cite{AMM}. By factoring out this parameter,
$\chi^{\alpha\beta}(v,v')$ becomes dimensionless. The most general
decomposition of this form factor involves two real, scalar functions
$\chi_2(y)$ and $\chi_3(y)$ defined by \cite{Luke}
\begin{equation}
   \chi^{\alpha\beta}(v,v')
   = (v'^\alpha\gamma^\beta - v'^\beta\gamma^\alpha)\,\chi_2(y)
   - 2 i\sigma^{\alpha\beta}\,\chi_3(y) .
\end{equation}
Irrespective of the structure of the current $J$, the form factor
$\chi_3(y)$ appears always in the following combination with $\xi(y)$:
\begin{equation}\label{combi}
   \xi(y) + 2 Z \bar\Lambda
   \bigg( {d_M\over m_Q} + {d_{M'}\over m_{Q'}} \bigg) \chi_3(y) \,,
\end{equation}
where $d_P=3$ for a pseudoscalar and $d_V=-1$ for a vector meson. It
thus effectively renormalizes the leading Isgur-Wise function,
preserving its normalization at $y=1$ since $\chi_3(1)=0$ according to
Luke's theorem \cite{Luke}. Eq.~(\ref{combi}) shows that knowledge of
$\chi_3(y)$ is needed if one wants to relate processes which are
connected by the spin symmetry, such as $B\to D\,\ell\,\nu$ and
$B\to D^*\ell\,\nu$.

Being hadronic form factors, the universal functions in HQET can only
be investigated using nonperturbative methods. QCD sum rules have
become very popular for this purpose. They have been reformulated in the
context of the effective theory and have been applied to the study of
meson decay constants and the Isgur-Wise functions both in leading and
next-to-leading order in the $1/m_Q$ expansion
\cite{Buch,Rady,Broa,SR,SR2,Sublea,Baga,Blok,MZY,2loop}. In particular,
it has been shown that very simple predictions for the spin-symmetry
violating form factors are obtained when terms of order $\alpha_s$ are
neglected, namely \cite{Sublea}
\begin{eqnarray}\label{leading}
   \chi_2(y) &=& 0 \,, \nonumber\\
   \chi_3(y) &\propto& \langle\bar q\,g_s\sigma_{\alpha\beta}
    G^{\alpha\beta} q\rangle\,\big[ 1 - \xi(y) \big] \,.
\end{eqnarray}
In this approach $\chi_3(y)$ is proportional to the mixed quark-gluon
condensate, and it was estimated that $\chi_3(y)\sim 1\%$ for large
recoil ($y\sim 1.5$). In a recent work we have refined the prediction
for $\chi_2(y)$ by including contributions of order $\alpha_s$ in the
sum rule analysis \cite{MZY}. We found that these are as important as
the contribution of the mixed condensate in (\ref{leading}). It is,
therefore, worthwhile to include such effects also in the analysis of
$\chi_3(y)$. This is the purpose of this article.

\section{Derivation of the sum rule}

The QCD sum rule analysis of the functions $\chi_2(y)$ and $\chi_3(y)$
is very similar. We shall, therefore, only briefly sketch the general
procedure and refer for details to Refs.~\cite{Sublea,MZY}. Our
starting point is the correlator
\begin{eqnarray}\label{correl}
   &&\int\!{\rm d}x\,{\rm d}x'\,{\rm d}z\,
    e^{i(k'\cdot x' - k\cdot x)}\,\langle 0\,|\,T\Big\{
    \big[\,\bar q\,\overline{\Gamma}_{M'}h'\,\big]_{x'},
    J(0), {\cal O}_{\rm mag}(z), \big[\,\bar h\,\Gamma_M\,q\,\big]_x
    \Big\}|\,0\,\rangle \nonumber\\
   &&= \Xi_2(\omega,\omega',y)\,{\rm tr}\Big\{
    (v'^\alpha\gamma^\beta - v'^\beta\gamma^\alpha)\,
    \overline{\Gamma}_{M'} P_+' \Gamma\,P_+\,i\sigma_{\alpha\beta}
    P_+ \Gamma_M \Big\} \nonumber\\
   &&\mbox{}\!+ \Xi_3(\omega,\omega',y)\,{\rm tr}\Big\{
    2 \sigma^{\alpha\beta}\,\overline{\Gamma}_{M'} P_+' \Gamma\,P_+
    \sigma_{\alpha\beta} P_+ \Gamma_M \Big\} \,,
\end{eqnarray}
where $P_+'=\case{1}/{2}(1+\rlap/v')$, and we omit the velocity labels
in $h$ and $h'$ for simplicity. The heavy-light currents interpolate
pseudoscalar or vector mesons, depending on the choice $\Gamma_M =
-\gamma_5$ or $\Gamma_M = \gamma_\mu-v_\mu$, respectively. The external
momenta $k$ and $k'$ in (\ref{correl}) are the ``residual'' off-shell
momenta of the heavy quarks. Due to the phase redefinition of the
effective heavy quark fields in HQET, they are related to the total
momenta $P$ and $P'$ by $k=P-m_Q\,v$ and $k'=P'-m_{Q'}\,v'$ \cite{Geor}.

The coefficient functions $\Xi_i$ are analytic in $\omega=2 v\cdot k$
and $\omega'=2 v'\cdot k'$, with discontinuities for positive values of
these variables. They can be saturated by intermediate states which
couple to the heavy-light currents. In particular, there is a
double-pole contribution from the ground-state mesons $M$ and $M'$.
To leading order in the $1/m_Q$ expansion the pole position is at
$\omega=\omega'=2\bar\Lambda$. In the case of $\Xi_2$, the residue of
the pole is proportional to the universal function $\chi_2(y)$. For
$\Xi_3$ the situation is more complicated, however, since insertions of
the chromo-magnetic operator not only renormalize the leading
Isgur-Wise function, but also the coupling of the heavy mesons to the
interpolating heavy-light currents ({\it i.e.}, the meson decay
constants) and the physical meson masses, which define the position of
the pole.\footnote{There are no such additional terms for $\Xi_2$
because of the peculiar trace structure associated with this
coefficient function.}
The correct expression for the pole contribution to $\Xi_3$ is
\cite{Sublea}
\begin{eqnarray}\label{pole}
   \Xi_3^{\rm pole}(\omega,\omega',y) &=&
    {F^2\over(\omega-2\bar\Lambda+i\epsilon)
             (\omega'-2\bar\Lambda+i\epsilon)} \nonumber\\
   &&\times\Bigg\{ \bar\Lambda\,\chi_3(y) + G_2\,\xi(y)
    + {\bar\Lambda\,\delta\Lambda_2\,\xi(y)\over
       (\omega-2\bar\Lambda+i\epsilon)} \Bigg\} \,.
\end{eqnarray}
Here $F$ is the analog of the meson decay constant in the effective
theory ($F\sim f_M\sqrt{m_M}$), $G_2$ is the spin-symmetry violating
correction to it, and $\delta\Lambda_2$ denotes the spin-symmetry
violating mass shift of the meson masses at order $1/m_Q$. More
precisely, these quantities are defined by \cite{SR2}
\begin{eqnarray}\label{FGLam}
   m_M - m_Q &=& \bar\Lambda \bigg\{1 + {d_M\over m_Q}\,\delta\Lambda_2
    + \ldots \bigg\} \,, \nonumber\\
   \langle 0\,|\,j(0)\,| M(v)\rangle &=& {i F\over 2}\,
    {\rm tr}\big\{\,\Gamma\,{\cal M}(v)\big\} \,, \\
   \langle 0\,|\,i\!\int\!{\rm d}z\,
   T\big\{j(0),{\cal O}_{\rm mag}(z)\big\}\,| M(v) \rangle
   &=& {i F\over 2}\,G_2\,{\rm tr}\Big\{2\sigma^{\alpha\beta}
    \Gamma\,P_+\sigma_{\alpha\beta}{\cal{M}}(v)\Big\} \,, \nonumber
\end{eqnarray}
where the ellipses represent spin-symmetry conserving or higher order
power corrections, and $j=\bar q\,\Gamma\,h(v)$. In terms of the
vector--pseudoscalar mass splitting, the parameter $\delta\Lambda_2$ is
given by $m_V^2-m_P^2= -8\bar\Lambda\,\delta\Lambda_2$.

For not too small, negative values of $\omega$ and $\omega'$, the
coefficient function $\Xi_3$ can be approximated as a perturbative
series in $\alpha_s$, supplemented by the leading power corrections in
$1/\omega$ and $1/\omega'$, which are proportional to vacuum
expectation values of local quark-gluon operators, the so-called
condensates \cite{SVZ}. This is how nonperturbative corrections are
incorporated in this approach. The idea of QCD sum rules is to match
this theoretical representation of $\Xi_3$ to the phenomenological pole
contribution given in (\ref{pole}). To this end, one first writes the
theoretical expression in terms of a double dispersion integral,
\begin{equation}
   \Xi_3^{\rm th}(\omega,\omega',y) = \int\!{\rm d}\nu\,{\rm d}\nu'\,
   {\rho_3^{\rm th}(\nu,\nu',y)\over(\nu-\omega-i\epsilon)\,
   (\nu'-\omega'-i\epsilon)} + {\rm subtractions} ,
\end{equation}
and performs a Borel transformation in $\omega$ and $\omega'$. This
yields an exponential damping factor, which suppresses the contamination
from higher resonance states and eliminates possible subtraction terms.
Because of the flavor symmetry it is natural to set the Borel
parameters associated with $\omega$ and $\omega'$ equal: $\tau=\tau'=
2 T$. One then introduces new variables $\omega_\pm=
\case{1}/{2}(\nu\pm\nu')$ and integrates over $\omega_-$. At this
stage, one employs quark-hadron duality to argue that the remaining
integral over the ``diagonal'' variable $\omega_+$ above a threshold
$\omega_0$ is dual to the contribution of higher resonance states
\cite{SR,Blok}. One thus equates the integral up to $\omega_0$ to the
Borel transform of the pole contribution in (\ref{pole}). This gives
the QCD sum rule
\begin{equation}\label{sumrul}
   \bigg\{ \bar\Lambda\,\chi_3(y) + \bigg[ G_2
   - {\bar\Lambda\,\delta\Lambda_2\over 2T} \bigg] \xi(y) \bigg\}
   F^2\,e^{-2\bar\Lambda/T} = \int\limits_0^{\omega_0}\!{\rm d}\omega_+
   \,e^{-\omega_+/T}\,\widetilde{\rho}_3^{\rm\,th}(\omega_+,y)
   \equiv K(T,\omega_0,y) \,.
\end{equation}
The effective spectral density $\widetilde{\rho}_3^{\rm\,th}$ arises
after integration of the double spectral density over $\omega_-$. Note
that for each contribution to it the dependence on $\omega_+$ is known
on dimensional grounds. It thus suffices to calculate directly the
Borel transform of the individual contributions to $\Xi_3^{\rm th}$,
corresponding to the limit $\omega_0\to\infty$ in (\ref{sumrul}). The
$\omega_0$-dependence can be recovered at the end of the calculation.

When terms of order $\alpha_s$ are neglected, contributions to the sum
rule for $\Xi_3$ can only be proportional to condensates involving the
gluon field, since there is no way to contract the gluon contained in
${\cal O}_{\rm mag}$. The leading power correction of this type is
represented by the diagram shown in Fig.~\ref{fig:1}(d). It is
proportional to the mixed quark-gluon condensate and, as shown in
Ref.~\cite{Sublea}, leads to (\ref{leading}). Here we are interested in
the additional contributions arising at order $\alpha_s$. They are
shown in Fig.~\ref{fig:1}(a)-(c). Besides a two-loop perturbative
contribution, one encounters further nonperturbative corrections
proportional to the quark and the gluon condensate.

Let us first present the result for the nonperturbative power
corrections. We find
\begin{equation}
   K_{\rm cond}(T,\omega_0,y) =
   {\alpha_s\langle\bar q q\rangle\,T\over 6\pi}\,
   \Big[ 2 - r(y) \Big]\,\delta_0\Big({\omega_0\over T}\Big)
   + {\langle\alpha_s GG\rangle\over 96\pi} \bigg({2\over y+1}\bigg)
   - {\langle\bar q\,g_s\sigma_{\alpha\beta}G^{\alpha\beta} q\rangle
      \over 48 T} \,,
\end{equation}
where
\begin{eqnarray}
   r(y) &=& {1\over\sqrt{y^2-1}}\,\ln\big(y + \sqrt{y^2-1}\big) \,,
    \nonumber\\
   \delta_n(x) &=& {1\over\Gamma(n+1)}\int\limits_0^x\!{\rm d}z\,
   z^n\,e^{-z} \,.
\end{eqnarray}
We think it is safe to truncate the series of power corrections after
the mixed condensate, since effects of higher dimension condensates are
expected to be very small. The next-order power corrections would come
from four-quark operators, which would contribute at the level of
$\vert\langle\bar q q\rangle\vert/T^3\sim 1-5\%$ as compared to the
quark condensate.

The calculation of the perturbative contribution is more cumbersome.
Two-loop diagrams such as that depicted in Fig.~\ref{fig:1}(a) were
considered in detail in Ref.~\cite{2loop}. We use dimensional
regularization and the integral representations discussed in this
reference. After Borel transformation we find
\begin{eqnarray}
   K_{\rm pert}(T,\infty,y) &=& 4 N_c C_F\,g_s^2\,\Gamma(D-1)\,
    {(2T)^{2D-4}\over(4\pi)^D} \nonumber\\
   &&\times \int\limits_0^1\!{\rm d}\lambda\,\lambda^{1-D}\!
    \int\limits_\lambda^\infty\!{\rm d}u_1
    \int\limits_{1/\lambda}^\infty\!{\rm d}u_2\,
    {(u_1 u_2-1)^{D/2-2}\over\big[ u_1+2(y+1)(u_2-1)\big]^{D-1}} \,,
\end{eqnarray}
where $C_F=(N_c^2-1)/2 N_c$, and $D$ is the dimension of space-time.
For $D=4$, the integrand diverges as $\lambda\to 0$. To regulate the
integral, we assume $D<2$ and use a triple integration by parts in
$\lambda$ to obtain an expression which can be analytically continued
to the vicinity of $D=4$. Next we set $D=4+2\epsilon$, expand in
$\epsilon$, write the result as an integral over $\omega_+$, and
introduce back the continuum threshold. This gives
\begin{eqnarray}\label{Kpert}
   K_{\rm pert}(T,\omega_0,y) &=& - {\alpha_s\over 48\pi^3}
    \bigg({2\over y+1}\bigg)^2 \int\limits_0^{\omega_0}\!
    {\rm d}\omega_+\,\omega_+^3\,e^{-\omega_+/T} \\
   &&\times \bigg\{ {1\over\epsilon} + 2\gamma_E - 2\ln 4\pi
    + 4\ln\omega_+ - y\,r(y) - 2\ln{y+1\over 2} - {23\over 6}
    + {\cal O}(\epsilon) \bigg\} \,. \nonumber
\end{eqnarray}
On first sight, the appearance of a divergence at order $\alpha_s$
seems surprising. Since the leading contribution to the spin-symmetry
violating form factors is of order $g_s$, one would expect divergences
to appear at order $g_s^3$. In fact, the one-loop renormalization group
equations which control the running of the hadronic quantities
$\chi_3(y)$ and $\delta\Lambda_2$ are homogeneous equations, confirming
this assertion. But $G_2$ satisfies the inhomogeneous equation
\cite{SR2}
\begin{equation}\label{RGE}
   \bigg\{ \mu{\partial\over\partial\mu} + {3\alpha_s\over 2\pi}
   \bigg\}\,G_2(\mu) = {2\alpha_s\over 9\pi}\,\bar\Lambda \,,
\end{equation}
which shows that divergences arise at order $\alpha_s$. At this order,
the renormalization of the sum rule is thus accomplished by a
renormalization of the ``bare'' parameter $G_2$ in (\ref{sumrul}). In
the $\overline{\rm MS}$ subtraction scheme, one defines a renormalized
parameter $G_2(\mu)$ by
\begin{equation}\label{G2mu}
   G_2(\mu) = G_2^{\rm bare} + {\alpha_s\over 9\pi}\,\bar\Lambda\,
   \bigg({1\over\epsilon} + \gamma_E - \ln{4\pi\over\mu^2}\bigg)
   + {\cal O}(g_s^3) \,.
\end{equation}
Hence a counterterm proportional to $\bar\Lambda\,\xi(y)$ has to be
added to the bracket on the left-hand side of the sum rule
(\ref{sumrul}). To evaluate its effect on the right-hand side, we note
that in $D$ dimensions \cite{Sublea}
\begin{eqnarray}\label{xilam}
   \bar\Lambda\,\xi(y)\,F^2\,e^{-2\bar\Lambda/T} &=&
    {3\over 16\pi^2} \bigg({2\over y+1}\bigg)^2
    \int\limits_0^{\omega_0}\!{\rm d}\omega_+\,\omega_+^3\,
    e^{-\omega_+/T} \\
   &&\times \bigg\{ 1 + \epsilon \bigg[ \gamma_E - \ln 4\pi
    + 2\ln\omega_+ - \ln{y+1\over 2} - 2 \bigg] \bigg\}
    + {\cal O}(g_s,\epsilon^2) \,. \nonumber
\end{eqnarray}
{}From (\ref{sumrul}), (\ref{Kpert}), (\ref{G2mu}) and (\ref{xilam}) it
is seen that indeed the $1/\epsilon$ pole cancels upon renormalization
of $G_2$. We find
\begin{eqnarray}\label{musum}
   &\bigg\{& \!\bar\Lambda\,\chi_3(y) + \bigg[ G_2(\mu)
    - {\bar\Lambda\,\delta\Lambda_2\over 2T} \bigg] \xi(y) \bigg\}
    F^2\,e^{-2\bar\Lambda/T} \nonumber\\
   &=& {\alpha_s\over 48\pi^3} \bigg({2\over y+1}\bigg)^2
    \int\limits_0^{\omega_0}\!{\rm d}\omega_+\,\omega_+^3\,
    e^{-\omega_+/T} \bigg\{2\ln{\mu\over\omega_+} + {17\over 6}
    + \bigg[ y\,r(y) - 1 + \ln{y+1\over 2} \bigg] \bigg\} \nonumber\\
   &&+ K_{\rm cond}(T,\omega_0,y) \,.
\end{eqnarray}
In this expression, all hadronic parameters are defined at the scale
$\mu$. However, we have only made the $\mu$-dependence of $G_2(\mu)$
explicit since this is what we are sensitive to in our analysis.

According to Luke's theorem, the universal function $\chi_3(y)$
vanishes at zero recoil \cite{Luke}. Evaluating (\ref{musum}) for
$y=1$, we thus obtain a sum rule for $G_2(\mu)$ and $\delta\Lambda_2$.
It reads
\begin{equation}
   \bigg[ G_2(\mu) - {\bar\Lambda\,\delta\Lambda_2\over 2T} \bigg]
    F^2\,e^{-2\bar\Lambda/T} = {\alpha_s\over 24\pi^3}
    \int\limits_0^{\omega_0}\!{\rm d}\omega_+\,\omega_+^3\,
    e^{-\omega_+/T} \bigg\{\ln{\mu\over\omega_+} + {17\over 12} \bigg\}
    + K_{\rm cond}(T,\omega_0,1) \,,
\end{equation}
where we have used that $r(1)=1$. Precisely this sum rule has been
derived previously, starting from a two-current correlator, in
Ref.~\cite{SR2}. This provides a nontrivial check of our calculation.
Using the fact that $\xi(y) = \big[2/(y+1)\big]^2 + {\cal O}(g_s)$
according to (\ref{xilam}), we find that the $\mu$-dependent terms
cancel out when we eliminate $G_2(\mu)$ and $\delta\Lambda_2$ from the
sum rule for $\chi_3(y)$.

Before we present our final result, there is one more effect which has
to be taken into account, namely a spin-symmetry violating correction
to the continuum threshold $\omega_0$. Since the chromo-magnetic
interaction changes the masses of the ground-state mesons
[cf.~(\ref{FGLam})], it also changes the masses of higher resonance
states. Expanding the physical threshold as
\begin{equation}
   \omega_{\rm phys} = \omega_0 \bigg\{1
   + {d_M\over m_Q}\,\delta\omega_2 + \ldots \bigg\} \,,
\end{equation}
we expect $\delta\omega_2\simeq\delta\Lambda_2\simeq -0.12$ GeV, where
the numerical value follows from the observed mass splitting between
$B^*$ and $B$. The appearance of such a shift is also in the spirit of
QCD sum rules. Assume that we had constructed a sum rule for the
combination of form factors given in (\ref{combi}). Then the presence
of $1/m_Q$ corrections would affect the stability of the sum rule for
the Isgur-Wise function $\xi(y)$ alone. The parameters that guaranteed
optimal stability for $\xi(y)$ would have to be readjusted by an amount
of order $1/m_Q$ in order to provide optimal stability of the sum rule
for the combination (\ref{combi}). From such a self-consistent
analysis, the parameters $\delta\Lambda_2$ and $\delta\omega_2$ have
been determined in Ref.~\cite{SR2}. One finds the vector--pseudoscalar
mass splitting in excellent agreement with experiment, and
$\delta\omega_2 = -(0.10\pm 0.02)$ GeV, which is just what we expected
on physical grounds.

The contribution of $\delta\omega_2$ to the sum rule for $\chi_3(y)$
has been calculated in Ref.~\cite{Sublea}. Including this term, we now
present our final result:
\begin{eqnarray}\label{chi3sum}
   \chi_3(y)\,\bar\Lambda\,F^2\,e^{-2\bar\Lambda/T} &=&
    {\alpha_s T^4\over 8\pi^3} \bigg({2\over y+1}\bigg)^2
    \bigg[ y\,r(y) - 1 + \ln{y+1\over 2} \bigg]\,
    \delta_3\Big({\omega_0\over T}\Big) \nonumber\\
   &&+ {3\,\delta\omega_2\over 32\pi^2}\,\omega_0^3\,e^{-\omega_0/T}
    \bigg[ \bigg({2\over y+1}\bigg)^2 - \xi(y) \bigg] \nonumber\\
   &&+ {\alpha_s\langle\bar q q\rangle\,T\over 6\pi}\,
    \Big[ 2 - r(y) - \xi(y) \Big]\,\delta_0\Big({\omega_0\over T}\Big)
    \\
   &&+ {\langle\alpha_s GG\rangle\over 96\pi}\,
    \bigg[ {2\over y+1} - \xi(y) \bigg]
    - {\langle\bar q\,g_s\sigma_{\alpha\beta}G^{\alpha\beta} q\rangle
       \over 48 T}\,\Big[ 1 - \xi(y) \Big] \,. \nonumber
\end{eqnarray}
It explicitly exhibits the fact that $\chi_3(1)=0$.

\section{Numerical Analysis}

Let us now turn to the evaluation of the sum rule (\ref{chi3sum}). For
the QCD parameters we take the standard values
\begin{eqnarray}
   \langle\bar q q\rangle &=& - (0.23\,{\rm GeV})^3 \,, \nonumber\\
   \langle\alpha_s GG\rangle &=& 0.04\,{\rm GeV^4} \,, \nonumber\\
   \langle\bar q\,g_s\sigma_{\alpha\beta}G^{\alpha\beta} q\rangle
   &=& m_0^2\,\langle\bar q q\rangle ~,~~
    m_0^2 = 0.8\,{\rm GeV^2} \,.
\end{eqnarray}
Furthermore, we use $\delta\omega_2=-0.1$ GeV from above, and
$\alpha_s/\pi=0.1$ corresponding to the scale $\mu=2\bar\Lambda
\simeq 1$ GeV, which is appropriate for evaluating radiative
corrections in the effective theory \cite{SR}. The sensitivity of our
results to changes in these parameters will be discussed below. The
dependence of the left-hand side of (\ref{chi3sum}) on $\bar\Lambda$
and $F$ can be eliminated by using a QCD sum rule for these parameters,
too. It reads \cite{SR2}
\begin{equation}\label{Fsum}
   \bar\Lambda\,F^2\,e^{-2\bar\Lambda/T}
   = {9 T^4\over 8\pi^2}\,\delta_3\Big({\omega_0\over T}\Big)
   - {\langle\bar q\,g_s\sigma_{\alpha\beta}G^{\alpha\beta} q\rangle
      \over 4 T} \,.
\end{equation}
Similarly, we calculate the Isgur-Wise function from $\xi(y) =
F(T,\omega_0,y)/F(T,\omega_0,1)$, where \cite{SR}
\begin{equation}\label{xisum}
   F(T,\omega_0,y) = {3 T^3\over 4\pi^2} \bigg({2\over y+1}\bigg)^2
   \delta_2\Big({\omega_0\over T}\Big) - \langle\bar q q\rangle
   + {(2y+1)\over 3}\,{\langle\bar q\,g_s\sigma_{\alpha\beta}
   G^{\alpha\beta} q\rangle\over 4 T^2} \,.
\end{equation}
Combining (\ref{chi3sum}), (\ref{Fsum}) and (\ref{xisum}), we obtain
$\chi_3(y)$ as a function of $\omega_0$ and $T$. These parameters can
be determined from the analysis of a QCD sum rule for the correlator of
two heavy-light currents in the effective theory \cite{SR2,Baga}. One
finds good stability for $\omega_0=2.0\pm 0.3$ GeV, and the consistency
of the theoretical calculation requires that the Borel parameter be in
the range $0.6<T<1.0$ GeV. It supports the self-consistency of the
approach that, as shown in Fig.~\ref{fig:2}, we find stability of the
sum rule (\ref{chi3sum}) in the same region of parameter space. Note
that it is in fact the $\delta\omega_2$-term that stabilizes the sum
rule. Without it there were no plateau.

Over the kinematic range accessible in semileptonic $B\to D^{(*)}\ell\,
\nu$ decays, we show in Fig.~\ref{fig:3}(a) the range of predictions
for $\chi_3(y)$ obtained for $1.7<\omega_0<2.3$ GeV and $0.7<T<1.2$
GeV. From this we estimate a relative uncertainty of $\sim\pm 25\%$,
which is mainly due to the uncertainty in the continuum threshold. It
is apparent that the form factor is small, not exceeding the level of
1\%.\footnote{When comparing our result to the function $\chi_3^{\rm
ren}(y)$ shown in Fig.~2 of Ref.~\cite{Sublea}, one has to include a
renormalization factor, which is approximately given by
$\alpha_s^{-1/3}\simeq 1.5$.}

Finally, we show in Fig.~\ref{fig:3}(b) the contributions of the
individual terms in the sum rule (\ref{chi3sum}). Due to the large
negative contribution proportional to the quark condensate, the terms
of order $\alpha_s$, which we have calculated in this paper, cancel
each other to a large extent. As a consequence, our final result for
$\chi_3(y)$ is not very different from that obtained neglecting these
terms \cite{Sublea}. This is, however, an accident. For instance, the
order-$\alpha_s$ corrections would enhance the sum rule prediction by a
factor of two if the $\langle\bar q q\rangle$-term had the opposite
sign. From this figure one can also deduce how changes in the values of
the vacuum condensates would affect the numerical results. As long as
one stays within the standard limits, the sensitivity to such changes
is in fact rather small. For instance, working with the larger value
$\langle\bar q q\rangle=-(0.26\,{\rm GeV})^3$, or varying $m_0^2$
between 0.6 and 1.0 GeV$^2$, changes $\chi_3(y)$ by no more than
$\pm 0.15\%$.

In conclusion, we have presented the complete order-$\alpha_s$ QCD sum
rule analysis of the subleading Isgur-Wise function $\chi_3(y)$,
including in particular the two-loop perturbative contribution. We find
that over the kinematic region accessible in semileptonic $B$ decays
this form factor is small, typically of the order of 1\%. When combined
with our previous analysis \cite{MZY}, which predicted similarly small
values for the universal function $\chi_2(y)$, these results strongly
indicate that power corrections in the heavy quark expansion which are
induced by the chromo-magnetic interaction between the gluon field and
the heavy quark spin are small.

\acknowledgements
It is a pleasure to thank Michael Peskin for helpful discussions. M.N.
gratefully acknowledges financial support from the BASF
Aktiengesellschaft and from the German National Scholarship Foundation.
Y.N. is an incumbent of the Ruth E. Recu Career Development chair, and
is supported in part by the Israel Commission for Basic Research and by
the Minerva Foundation. This work was also supported by the Department
of Energy, contract DE-AC03-76SF00515.

\figure{\label{fig:1}
Diagrams contributing to the sum rule for the universal form factor
$\chi_3(v\cdot v')$: two-loop perturbative contribution (a), and
nonperturbative contributions proportional to the quark condensate (b),
the gluon condensate (c), and the mixed condensate (d). Heavy quark
propagators are drawn as double lines. The square represents the
chromo-magnetic operator.}

\figure{\label{fig:2}
Analysis of the stability region for the sum rule (\ref{chi3sum}): The
form factor $\chi_3(y)$ is shown for $y=1.5$ as a function of the Borel
parameter. From top to bottom, the solid curves refer to
$\omega_0=1.7$, 2.0, and 2.3 GeV. The dashes lines are obtained by
neglecting the contribution proportional to $\delta\omega_2$.}

\figure{\label{fig:3}
(a) Prediction for the form factor $\chi_3(v\cdot v')$ in the stability
region $1.7<\omega_0<2.3$ GeV and $0.7<T<1.2$ GeV. (b) Individual
contributions to $\chi_3(v\cdot v')$ for $T=0.8$ GeV and $\omega_0=2.0$
GeV: total (solid), mixed condensate (dashed-dotted), gluon condensate
(wide dots), quark condensate (dashes). The perturbative contribution
and the $\delta\omega_2$-term are indistinguishable in this figure and
are both represented by the narrow dots.}

\end{document}